\documentclass[a4paper, 11pt]{article}
\usepackage{graphicx}
\usepackage{siunitx}
\DeclareSIUnit\Molar{\textsc{m}}
\usepackage[colorlinks=true, linkcolor=blue, urlcolor=blue, 
citecolor=blue, linktocpage=true]{hyperref}

\title{Analysis of claims that the brain extracellular impedance is high and non-resistive}
\author{Boris Barbour\\Institut de Biologie de l'Ecole Normale
    Sup\'erieure (IBENS),\\ Ecole Normale Sup\'erieure, CNRS UMR 8197,
    INSERM U1024,\\ PSL Research University, Paris, F-75005,
    France.}

\begin{document}

\maketitle
\begin{abstract}{\bf
Numerous measurements in the brain of the impedance between two extracellular electrodes have shown that it is approximately resistive in the range of biological interest, $<\SI{10}{\kilo\hertz}$, and has a value close to that expected from the conductivity of physiological saline and the extracellular volume fraction in brain tissue. Recent work from the group of Claude B\'edard and Alain Destexhe has claimed that the impedance of the extracellular space is some three orders of magnitude greater than these values and also displays a $1/\sqrt{f}$ frequency dependence (above a low-frequency corner frequency). Their measurements were performed between an \emph{intracellular} electrode and an extracellular electrode. It is argued that they incorrectly extracted the extracellular impedance because of an inaccurate representation of the large, confounding impedance of the neuronal membrane. In conclusion, no compelling evidence has been provided to undermine the well established and physically plausible consensus that the brain extracellular impedance is low and approximately resistive.}
\end{abstract}

This analysis will focus on the most recent article of the B\'edard and Destexhe group \cite{Gomes2016}.

There have been numerous measurements of the impedance between two extracellular electrodes in the brain. In this configuration, current flow may cross cell membranes, but is not constrained to do so. Perhaps a little surprisingly given the large intracellular volume bounded by capacitive membranes, the consensus from these measurements is that the extracellular impedance is approximately resistive up to quite high frequencies of about \SI{10}{\kilo\hertz}, encompassing the entire range of interest for biological signals. The conductivity in this range below \SI{10}{\kilo\hertz} is about \SIrange{0.3}{0.6}{\siemens m^{-1}} \cite{Ranck1963,Nicholson1975,Okada1994,Logothetis2007,Goto2010}. These studies have recently been reviewed in \cite{Miceli2017}. It should be noted that these values are of the order of magnitude expected for the extracellular space alone (indeed somewhat lower), given the conductivity of physiological saline (about \SI{1.7}{\siemens m^{-1}}) and the volume fraction (0.2) of brain tissue \cite{Gardner-Medwin1980,Syková2008}.

Here, instead of measuring the extracellular impedance between extracellular electrodes, Gomes et al.~perform their measurement between intracellular and extracellular electrodes, which forces current flow to cross the cell membrane. This configuration therefore introduces the confounding and elevated impedances of the recording electrode, cell and cell membrane in series with the extracellular impedance, but Gomes et al.~justify this by the inclusion of the `natural interface' between the membrane and the extracellular fluid, which they argue is an important feature that has hitherto been neglected. They report that the extracellular impedance extracted from these measurements is not resistive but displays a $1/\sqrt{f}$ dependence (above a low-frequency corner frequency), where $f$ is the frequency. This frequency dependence is attributed to a Warburg impedance, a concept borrowed from electrochemistry. A Warburg impedance arises when reactive species are depleted or accumulated at the electrode interface by the current flow in solution carried by those same species. The Warburg impedance is frequency dependent because the depletion/accumulation takes time.

A first issue is semantic, but important. Any special properties of the membrane-solution interface should be considered to be part of the membrane impedance and not as part of the extracellular impedance.

A diagram of the measurement setup is shown in Fig.~1 (of Gomes et al.). This seems to contradict the statement that `the reference electrode is passive, just measuring the extracellular voltage' (Methods). As drawn, any current circulating in the recording electrode must also traverse the reference electrode, so it would not be `passive'. Moreover, Gomes et al.~state that `variations on the extracellular electrode did not exceed \SI{1}{\%} of the variations of the intracellular potential'. However, no circuit is shown or described that is able to measure the extracellular voltage independently.

It is useful to quantify the expected extracellular impedance. Assume the soma of a neurone is a sphere of radius $r = \SI{5}{\micro m}$. The resistance of the extracellular space from the spherical surface to infinity is given by
\begin{equation}R = \frac{1}{4 \pi r \sigma},\end{equation}
where $\sigma = \SI{0.3}{\siemens m^{-1}}$ is the conductivity of the extracellular space. We obtain $R \approx \SI{50}{\kilo\ohm}$. This is the extracellular impedance that one would expect to measure using the cell soma as an electrode, according to the standard values of extracellular conductivity. This value represents an upper bound, because if any part of the dendritic surface also contributed current flow, the effective radius would be larger and the resistance even lower. This value is to be compared with an electrode resistance of tens of \SI{}{\mega\ohm} and a membrane resistance greater than \SI{100}{\mega\ohm} at low frequencies. 

The potential difficulty of accurately extracting the extracellular im\-ped\-ance from the series combination with the much higher electrode and neuronal impendances is apparent. Departing temporarily from the nomenclature of Gomes et al., they measure a global impedance
\begin{equation}Z_\mathrm{global} = Z_\mathrm{electrode} + Z_\mathrm{neurone} + Z_\mathrm{extracellular}\end{equation}
To extract $Z_\mathrm{extracellular}$, Gomes et al.~need to subtract $Z_\mathrm{electrode}$ and $Z_\mathrm{neurone}$ from the sum. Because these confounding impedances are likely to be much greater than $Z_\mathrm{extracellular}$, they must be determined with a precision of better than 1 part in 10000. The article contains no independent method at all for determining $Z_\mathrm{electrode}$ or $Z_\mathrm{neurone}$, let alone to such precision. Instead, all components are extracted from a single fit, whose ability to identify each component securely is not established.

Gomes et al.~represent the impedance of the neurone's dendritic tree using an approximation they call the `generalised cable formalism'. The impedance of the dendritic tree is expected to dominate the global impedance measured, so the failure of the generalised cable formalism to account entirely for the measured impedance potentially arises from a limitation of the model representation or a fitting problem (see below). However, instead of adjusting their dendrite model, Gomes et al.~obtain a better fit by the ad hoc inclusion of a series impedance element with a $1/\sqrt{f}$ dependence. This is justified as arising from a Warburg impedance. However, at least two studies have shown that intracellular impedance spectra of pyramidal cells very similar to that studied here can be fully accounted for by a multi-compartment neuronal model with a negligible extracellular resistance \cite{Yaron-Jakoubovitch2008,Miceli2017}.

The extracellular Warburg impedance proposed by Gomes et al.~is given by their Eqn.~(2) and at low frequencies approaches the limit of $\sqrt{A^2 + B^2}$, where parameters $A$ and $B$ were obtained by fitting and are given in the legend of their Fig.~2 as $A = \SI{99}{\mega\ohm}$ and $B = \SI{3.8}{\mega\ohm}$. The impedance in the low-frequency limit is therefore about \SI{99}{\mega\ohm}. This is some 2000-fold greater than the \SI{50}{\kilo\ohm} expected from previous measurements. Even at higher frequencies, the Warburg impedance only falls off quite slowly, because of the $1/\sqrt{f}$ dependence above the corner frequency of $f_W = \SI{36}{\hertz}$. At a typical signal frequency for an action potential of \SI{1}{\kilo\hertz}, the Warburg impedance would only have fallen to around \SI{20}{\%} of its low-frequency value, about \SI{20}{\mega\ohm}. Such high values appear incompatible with everyday experience of intra- and extracellular recordings. Thus, one can routinely record intracellular sodium currents of nanoAmp\`eres from somata of the size assumed here. Yet each nanoamp flowing across an extracellular impedance of \SI{20}{\mega\ohm} would generate an extracellular signal of \SI{20}{mV} in amplitude. To my knowledge extracellular signals of this amplitude have not previously been reported. This issue can be put another way: the extracellular impedance reported by Gomes et al.~is of the same order of magnitude as a membrane impedance. If this were true, one would  expect voltage responses to a given current to be of similar amplitude in the extracellular space as across the membrane.

There are possible problems with the authors' inclusion of this Warburg impedance. Normally the mechanism of the Warburg impedance requires depletion/accumulation of charges that contribute to electrode reactions, but there is no oxidation or reduction occurring at the membrane. Gomes et al.~do not establish an equivalence with ions that do cross the membrane, presumably mostly potassium ions, although this may be their implicit assumption. There is no attempt to model the changes of potassium ion concentrations and the resulting change of extracellular conductivity. In any case, for sufficiently small currents the effects of this mechanism would disappear, because the concentration changes would become small relative to the resting ionic concentrations. Instead of constraining their model Warburg impedance with available information, Gomes et al.~introduce it with three free parameters, enabling it to attain unrealistically high impedances during the fitting procedure.

A further problem is that ionic currents only account for a significant fraction of the membrane current at low frequencies. Thus, above about 10 Hz, a majority of the membrane current is capacitive and that fraction increases further with frequency. Because capacitive currents only involve displacement of ions relative to the membrane, no ionic transfer and no electrode reaction are required; this weakens the authors' argument that there is a special property of the membrane-solution interface that has not been accurately captured in previous measurements.

The model fitting appears to have been difficult. According to the methods, `traditional fitting methods' did not work, so the parameter space was probed by a Monte Carlo method. However, the density of sampling seems to have been rather low: at most 5000 samples from `very large' domains. Some of the models had 6 parameters, so this amounts to an effective grid of little more than 4 (sixth root of 5000) random values per dimension. In turn, imprecision of fitting might undermine the model evaluation.

In their most recent preprint, the B\'edard and Destexhe group suggest that previous measurements of the extracellular space might have been affected by tissue damage from the electrode \cite{2016arXiv161110047B}. It is unlikely that this mechanism is quantitatively important or that it could account for the difference between the two sets of measurements. Note first that the measurements of the B\'edard and Destexhe group are likely to have quite similar tissue damage, as two electrodes are still inserted into the tissue. The mechanism therefore cannot explain the difference between their and previous measurements.

\begin{figure}
\begin{center}
\includegraphics[scale=0.7857]{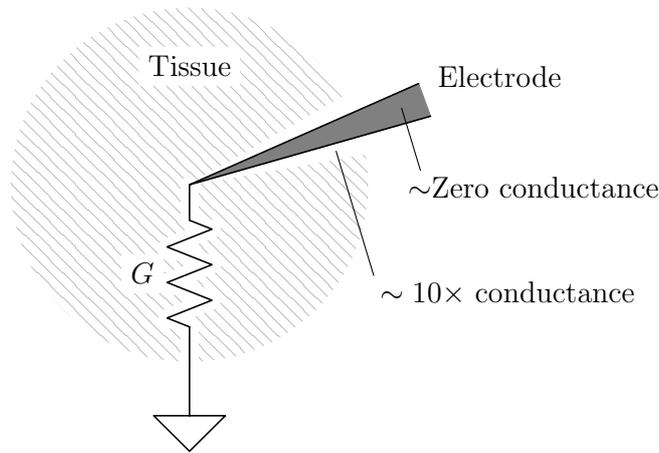}
\caption{Diagram for understanding the effect of an electrode on the measurement of the extracellular impedance $G$. A thin layer around the electrode will exhibit increased conductance, while the electrode itself will have essentially zero conductance, by virtue of its connection to a high-impedance amplifier. For an electrode-affected cone with a generous plane angle of \ang{20} (as drawn), the solid angle it subtends at the electrode would be 0.1 steradians or less than 1\,\% of $4\pi$.}
\label{electrode}
\end{center}
\end{figure}

Insertion of an electrode will have two opposing effects on the tissue resistance (Fig.~\ref{electrode}). If tissue is damaged or cleared from a space around the electrode, that will indeed increase the conductivity somewhat, compared to normal tissue. Essentially, the tortuosity of the extracellular space would decrease and its volume fraction would increase \cite{Gardner-Medwin1980,Syková2008}; the combined effect of these two alterations could not exceed a roughly ten-fold increase of conductivity and would apply only to a small fraction of the tissue. Conversely, the electrode itself would represent a volume with an effectively zero conductivity, because it is connected to an amplifier with a very high impedance. It is therefore unclear whether an electrode would increase or decrease tissue impedance. In any case, because the the solid angle subtended by any electrode as viewed from its tip is likely to be quite small, the overall change of resistance between the tissue at the electrode tip and infinity is likely to be small.

In conclusion, the observations attributed by Gomes et al.~to the extracellular impedance may instead arise from an incorrect representation of the membrane impedance of the complex dendritic tree, which confounds extraction of the extracellular impedance from an extremely difficult measurement configuration. This in turn would imply that there is no evidence calling into question the previous determinations of a low and largely resistive extracellular impedance in brain tissue. Similar arguments may be relevant to other publications on this subject by the same group \cite{Bedard2004,Bedard2010}.\footnote{Readers may find further discussion of this subject 
on PubMed Commons:\\ \url{https://www.ncbi.nlm.nih.gov/pubmed/26745426}}


\bibliographystyle{plain}
\bibliography{destexhe}

\end{document}